\documentstyle [12pt]{article}
\newcommand{\delbw}{\mbox{$\stackrel{\leftrightarrow}{\Delta}$}}
\newcommand{\delr}{\mbox{$\stackrel{\rightarrow}{\Delta}$}}
\newcommand{\dell}{\mbox{$\stackrel{\leftarrow}{\Delta}$}}
\begin{document}

\begin{titlepage}

\begin{flushright}
DAMTP-92-70
\end{flushright}

\vspace*{5mm}

\begin{center}
{\Huge Non-relativistic QCD for Heavy Quark Systems}\\[15mm]
{\large\it UKQCD Collaboration}\\[3mm]

{\bf S.M.~Catterall, F.R.~Devlin, I.T.~Drummond,
 R.R.~Horgan}\\
DAMTP, Silver St., University of Cambridge, Cambridge CB3~9EW, UK

{\bf A.D.~Simpson}\\
Department of Physics, The University of Edinburgh, Edinburgh EH9~3JZ,
Scotland

\end{center}
\vspace{5mm}

\begin{abstract}

We employ a nonrelativistic version of QCD (NRQCD) to study heavy
$q\bar{q}$-bound states
in the lowest approximation without fine structure. We use gluon configurations
on a $16^3\times 48$-lattice at $\beta=6.2$ from the UKQCD collaboration.
For a bare quark mass near that of the b-quark ($Ma=1.6$) we obtain the bound
state masses for the $S$, $P$ and both types of $D$-waves. We also detect
signals for two types of hybrids ($q\bar{q}g$-states). The results are
sufficiently accurate to confirm that the values of the $D$-wave mass from
both $D$-waves coincide thus indicating that the cubical invariance of the
lattice is restored to full rotational invariance to a good approximation.

We also study $S$ and $P$-wave masses for values of the range of bare
quark mass $Ma=1.0,~1.3,~1.6~\&~1.9$~.
The results confirm the idea that the $S/P$-splitting is relatively insensitive
to the value of the bare quark mass.

\end{abstract}
\vfill
\end{titlepage}

\section{Introduction}

Heavy quark bound states ($J/\psi, \Upsilon$) are not only intrinsically
interesting
but because of the nearly non-relativistic motion of the constituent quarks,
offer
the prospect of numerical calculations considerably more tractable than the
standard calculations for light quarks
\cite{eicht1,eicht2,beth,khadra,degrand}.
Several theoretical proposals have been made \cite{pet1}
and a number of numerical calculations carried out \cite {chris1,chris2}.

For quark masses $M \gg \Lambda_{QCD}$,
the quark velocity $v\ll 1$, so that its dynamics are well modelled
by an effective field theory, nonrelativistic QCD (NRQCD), in which
the kinetic energy term is nonrelativistic \cite{beth,pet1}.
Clearly this approximation will only be valid for momenta very much less
than some cut-off $\Lambda=a^{-1}$ ($a$ is the lattice
spacing) of the order of the heavy quark mass $M$. Furthermore the
effective action will contain correction terms which serve both
to implement relativistic effects such as the hyperfine structure splitting
and to compensate for the effects of the high momentum states which have been
removed from the theory. These may be reorganised into a power
series in the characteristic quark velocity determined by the dynamics.
In practice this series is truncated at some finite order in the quark
velocity and the couplings tuned so as to reproduce the low
energy behaviour of full QCD.

These systems serve as an important testing ground for lattice QCD
since finite volume effects are expected to be much less important
than for light quark states.
Furthermore the NRQCD approach avoids the problem of fermion
doubling and allows a rapid calculation of the quark propagator
as an initial value problem (since it is determined by
a lattice difference equation which is first order in time) \cite{beth,pet1}.
Issues to be decided
concern the appropriate size of lattice and the optimum values for the
couplings and masses
in the ``non-relativistic'' hamiltonian in order to best describe the bound
state.

In this letter we extend the numerical study of heavy quark bound
states in QCD to a higher value of the coupling constant $\beta=6.2$, using QCD
gauge field configurations on $16^3\times48$-lattice from the UKQCD
collaboration.
We have employed the lowest order approximation in which we retain
only the kinetic energy term in the effective action and studied
correlation functions corresponding to the ground state ($S$-wave) and angular
excitations ($P~\&~D$) of the
$q\bar{q}$-system. We have also looked for hybrid $q\bar{q}g$-states and
obtained a clear
though still noisy signal. We have examined the $S~\&~P$-waves for other
(lower) bare
masses for the quarks and detected trends in the bound state energies and level
splittings consistent
with the results of Davis and Thacker \cite{chris2}.

The important problem of fine structure of the energy levels we have not
tackled in this paper partly because of restrictions in computational
facilities. We intend to
explore these and other questions in future work.

\vfill
\pagebreak

\section{Propagators and Operators}

The evolution equation for the heavy quark propagator takes the form
\begin{equation}
G\left(x,t+1\right)=U^{\dagger}_t\left(x,t\right)
\left(\left(1-\frac{H_0}{n}\right)^n
G\left(x,t\right)\right)
+\delta\left(x\right)\delta\left(t\right)
\end{equation}
where $x$ labels the spatial position and $G\left(t\right)=0$ for $t\le 0$
and $n$ is the order of the update as discussed by Thacker and Davies
\cite{chris1}.
The modified update is necessary for stability at certain values of the bare
quark mass.
We have used $n=2~\&~3$, as appropriate. In fact there appeared to be little
sensitivity
to the value of $n$ in the results when sufficient stability was present.
The kinetic operator $H_0$ is defined as
\begin{equation}
H_0={-1\over 2Ma}\sum_{i=1}^3 \Delta^{+}_i\Delta^{-}_i
\end{equation}
The covariant finite differences $\Delta^{+}$, $\Delta^{-}$ are given by
their usual expressions (we have suppressed all colour and spin
indices)
\begin{equation}
\Delta^+_iG\left(x,t\right)=U_i\left(x,t\right)G\left(x+i,t\right)-G\left(
x\right)
\end{equation}
\begin{equation}
\Delta^-_iG\left(x,t\right)=G\left(x,t\right)-U^{\dagger}_i\left(x-i,t\right)
G\left(x-i,t\right)
\end{equation}
The correlation functions that we have considered are of the form
\begin{equation}
g_i\left(x,t\right)=\left\langle O_i\left(x,t\right)O^\dagger_i\left(0,0\right)
\right\rangle
\end{equation}
where the index i runs over the different lattice operators
corresponding to the $S$, $P$, $D$ and hybrid states.
For the $S$ wave $O_S\left(x,t\right)$ with quantum numbers $J^{PC}=0^{-+}$
\begin{equation}
O_S\left(x,t\right)=\chi^\dagger\left(x,t\right)\psi\left(x,t\right)
\end{equation}
The fields $\psi\left(x,t\right)$ and $\chi\left(x,t\right)$ represent the
quark and antiquark with propagators $G\left(x,t\right)$ and
$G^\dagger\left(x,t\right)$ respectively. In the current approximation
these are spin independent.
The higher angular momentum
states involve the correlations of two displaced quarks, the $P$ operator
with $J^{PC}=1^{+-}$ being
\begin{eqnarray}
O_P\left(x,t\right) & = &{1\over 2}\chi^\dagger\left(x,t\right)\delbw_i
			 \psi\left(x,t\right)\\
                   & = &{1\over 2}\left(\chi^\dagger\left(x,t\right)\left(
			\Delta_i \psi\left(x,t\right)\right)-\left(\Delta\chi
			\left(x,t\right)\right)^\dagger\psi\left(x,t\right)
			\right)
\end{eqnarray}
The choice of $\delbw$ (with $\Delta$ now
a forward derivative) ensures that the state is orthogonal to the
S state and yields a correlation function that is
entirely real. The continuum $D$ state
splits into two states under the lattice group, call them $D_1$ and $D_2$.
These correspond to the operators
\begin{equation}
O_{D_1}\left(x,t\right)={1\over 4}\chi^\dagger\left(x,t\right)\left\lbrace
\delbw_i\delbw_j\right\rbrace\psi\left(x,t\right)
\end{equation}
\begin{equation}
O_{D_2}\left(x,t\right)={1\over 4}\chi^\dagger\left(x,t\right)\left(
\delbw^2_i-\delbw^2_j\right)\psi\left(x,t\right)
\end{equation}
The indices $i$ and $j$ correspond to two (distinct) spatial directions
and the curly braces in $D_1$ indicate a symmetrised combination is to be
taken. The hybrid states we investigate fit conveniently into the same
computational
framework. They correspond to taking {\em commutators}
of the lattice covariant derivatives and therefore are only non-trivial in the
presence of a gauge field configuration. We have
\begin{equation}
O_{H_1}\left(x,t\right)={1\over 2}\chi^\dagger\left(x,t\right)\left(
\left[\delr_i,\delr_j\right]+\left[\dell_i,\dell_j\right]\right)\psi\left
(x,t\right)
\end{equation}
\begin{equation}
O_{H_2}\left(x,t\right)={1\over 2}\chi^\dagger\left(x,t\right)\left(
\left[\delr_i,\delr_j\right]-\left[\dell_i,\dell_j\right]\right)\psi\left
(x,t\right)
\end{equation}
These states possess quantum numbers $J^{PC}$ equal to $1^{-+}$
and $1^{--}$ respectively. Because we have not included fine structure
splitting in our hamiltonian the $1^{--}$ state does not mix with the
$q\bar{q}$-state
with the same quantum numbers.
An application of Wick's theorem then allows all the correlation
functions to be built up out of sums of (gauge invariant) loops on the
lattice constructed from the quark propagators and appropriate
link matrices.

\section{Results}

Our results are obtained from ten uncorrelated configurations
on a $16^3\times 48$ quenched lattice
at $\beta=6.2$~. They were used to generate the NRQCD correlation functions
with bare quark masses $Ma=1.0,~1.3,~1.6~\&~1.9$. The  configurations were
obtained using the 64 i860-node Meiko Computing Surface at Edinburgh.
The update consisted of a cycle of 1 three-subgroup Cabibbo-Marinari
heat-bath sweep followed by 5 over-relaxed sweeps.

In order to extract the
maximum information from each configuration
we used three starting timeslices spaced by $16$ units
in the temporal direction and evolved the propagators from
$4^3$ evenly spaced spatial starting points within each of these slices. In
addition
we performed an average over spatial directions for the higher angular
momentum states. It was found that each of these procedures contributed
to a reduction in  the
raggedness of the results from a particular configuration. Indeed the
ground state $S$-wave correlation function computed from a single
configuration was already rather smooth even out to 48 time slices.
The higher waves of course required more data.
A total of $6$ quark propagators were needed to reconstruct
all the correlators of interest. The errors were assessed by
treating the data from each configuration as statistically independent.

To extract estimates for the masses of the states we utilised a
two exponential fit with four parameters to the zero spatial
momentum correlator for times $t\le 24$.
\begin{equation}
G(t)=Ae^{-M_0 t}\left(1+Be^{-M_1 t}\right)~~.
\end{equation}
The basis for the fit was a simple least squares procedure based on the
statistical errors estimated from the data. The fitting procedure was
carried out for a sequence of initial time slices.
The numbers quoted are those obtained from the best fit consistent within
one standard deviation with those from succeeding initial time slices.
We further checked that
fits to a single exponential at large times gave statistically
consistent results.

For the case $Ma=1.6$
we used all 10 gauge field configurations.
If we use a value for the lattice spacing extracted from string tension
measurements, $a^{-1}=2.7$ Gev, to convert masses
to physical units this corresponds to $M=4.32$, roughly the b-quark mass.
The procedure worked well particularly for $S$ and $P$-waves.
The fits are shown in Fig. 1. Clearly for
these states we are in good quantitative control of both the statistical
and systematic errors. As a further check we
fitted the $S$-wave data out to $48$ timeslices and found a stable
plateau in the ground state mass.
Figs. 2a$~\&~$2b show the results for the two
$D$ states. Here we used the two exponential fit only for $t<20$ since this
yielded a more stable mass estimate. As we can see from Table 1. the
two $D$-state masses are equal to within the errors which is
an indication of the restoration of full rotational invariance at
large distance.

The data and fits for the hybrid states are plotted in Figs. 3a$~\&~$3b~.
Clearly
our data is much noisier for these states and we attempted only a rough single
exponential
fit for an intermediate range of time-slices. This yielded estimates for the
masses in the vicinity
of the $D$-wave mass for both hybrids (see Table 1.). Although the results are
very much
provisional, it is nevertheless encouraging to see a signal at this stage.
Hybrids
are of particular interest because they can only exist as a result of
propagation effects
in the gauge fields and cannot be realised directly in a simple quark potential
model.
We hope in the future to achieve better results for these and the more
conventional states.
This will require both higher statistics and the use
of improved operators with a stronger overlap on the physical
states.

On converting our lattice value for the $S/P$-splitting
to physical units we obtain $\Delta E_{SP}=0.35(2)$ Gev which is
quite close to the experimental number of $0.4$ for the splitting between
triplet $S$ and $P$-states of the $\Upsilon$-system.
It is also close to the implied physical value for the
splitting found in ref \cite{chris2}. This supports the physical ideas
underlying the method. We have further strengthened the basis for the
calculation
by investigating the
dependence of $S/P$-splitting on the bare quark mass with results at
$Ma=1.0,~1.3,1.6~\&~1.9$
summarised in Table 2. These confirm the approximate independence of
this quantity on the bare quark mass. It should be noted however that these
results which were
obtained for reasons of computational economy from only 3 configurations,
required the use of a third order update method in order to encompass the
lowest mass $Ma=1$~. These new results for $Ma=1.6$ using a third order update
are consistent with
the original results using a second order method at this value of the bare
quark mass.

Finally in Figs. 4a$~\&~$4b, we show effective mass plots for the $S$ and
$P$-waves
at $Ma=1.6$ and in Figs. 5a$~\&~$5b the same for $Ma=1.0$~. In both cases the
effective mass for the $S$-wave
shows a very convincing plateau. The $P$-wave at $Ma=1.6$ is rather convincing
also while the $P$-wave at $Ma=1.0$ is much noisier. It is a matter of
judgement whether these results can be taken as exhibiting the oscillations
for the effective mass encountered by Davies and Thacker \cite{chris2}. It
should be noted
however that our computed errors are somewhat larger than those claimed in ref
\cite{chris2}. The greater
noise in the results at the lower mass is  we suspect, due to the fact that
with a third order update the limit of stability is roughly $Ma=1.0$
\cite{chris1}~.
This potential instability underlines the need for using higher order updates.
It is therefore encouraging that when there is stability ($Ma=1.3,~1.6~\&~1.9$)
the results appear to
be insensitive to the order of the update. This is an important point that
should be more thoroughly checked.

\section{Conclusions}

We have investigated the application of the NRQCD method in a numerical
simulation
of the lowest order kinetic energy hamiltonian, thus neglecting fine structure
splitting for the purposes of this paper. We found that the $S/P$ splitting
was roughly independent of {\em a}) the bare mass of the quark
and {\em b}) the order of the time update used to compute the quark Green's
functions.
This encourages belief in the the physical meaning of the method. At the
highest mass value
we obtained results sufficiently accurate to confirm that the two versions of
the
$D$-wave were degenerate. This was true not only of the $D$-wave masses
themselves
but also of the additional parameters in the two exponential fit that mimic the
higher mass states that obviously affect the form of the Green's function at
small time separation.
The $S/D$-split is approximately twice that of the $S/P$-wave split.

Hybrid states in the quark model have long been of interest \cite{Close}. It is
encouraging therefore that were able to obtain
a signal for hybrid $q\bar{q}g$-states. Although the signal was noisy it
suggests that they
exist at a mass comparable to that of the $D$-wave. This implies an excitation
energy above
the ground state of $\sim$ .8 Gev, a value rather less than but not grossly out
of line with
that suggested by excited static potential calculations applied to the
$\Upsilon$ system
\cite{Mich1,Mich2}.

Calculations to check the dependence
of physical results on the bare quark mass suggests that this is weak as is the
dependence on the
order of the update. The absolute values of the energies of the states
investigated do depend on the
value of the bare quark mass.
We note a trend of {\em decreasing} $S$-wave mass with {\em increasing} quark
mass revealed in
Table 2. In fact our results suggest that $M_S^{-1}$ is approximately linear in
the
bare quark mass $M$ as one might intuitively expect. This is consistent
with the results of ref \cite{chris2}~. However we have not explored the
effects of mass renormalization
in this context, important though that issue is.

We believe that we have obtained encouraging results for the NRQCD method and
that it is possible
to extract good results from only a few configurations. We intend to extend our
calculations to
other sizes of lattice and other values of the coupling constant. This is
particularly
important since the lattice value of $\sim .13$ for the $S/P$-splitting
suggests $\Upsilon$ radii
of $\sim 6-8$ lattice units at $\beta=6.2$~. This is consistent with the
wavefunction radius
measured in ref\cite{chris2} and predicted in other calculations \cite{ronqm}~.
In order to
ensure that finite size effects do not influence our results at this value of
$\beta$ it is
important to repeat our calculations on a larger lattice. However because we
see reasonable agreement
with the results of ref\cite{chris2} obtained with $\beta=6.0$ on our size of
lattice we do not
expect there to be a strong effect.
Further improvements
in extracting data from QCD configurations can be envisaged such as better
statistical treatment
of the data and the use of better wave functions in the measurement of the more
difficult
states as well as higher statistics from more configurations.

\vfill
\pagebreak

\section*{Acknowledgements}

F R Devlin wishes to acknowledge the support of the Department of Education
of Northern Ireland.

\vfill
\pagebreak

\vfill
\pagebreak

\section*{Table Captions}
\begin{enumerate}
\item Masses for the states measured with $Ma=1.6$~using the second order
update.

\item The $M_S, M_P ~\&~S/P$-splitting at various masses using the third order
update.

\end{enumerate}

\section*{Figure Captions}
\begin{enumerate}

\item[] Fig. 1 $S$ and $P$ wave correlation functions at $Ma=1.6$ with solid
line fit (two exponential).

\item[] Fig. 2a $D_1$ wave correlation functions at $Ma=1.6$ with solid line
fit (two exponential).

\item[] Fig. 2b $D_2$ wave correlation functions at $Ma=1.6$ with solid line
fit (two exponential).

\item[] Fig. 3a $H_1$ hybrid correlation functions at $Ma=1.6$ with solid line
fit (single exponential).

\item[] Fig. 3b $H_2$ hybrid correlation functions at $Ma=1.6$ with solid line
fit (single exponential).

\item[] Fig. 4a Effective mass plot for $S$-wave at $Ma=1.6$~.

\item[] Fig. 4b Effective mass plot for $P$-wave at $Ma=1.6$~.

\item[] Fig. 5a Effective mass plot for $S$-wave at $Ma=1.0$~.

\item[] Fig. 5b Effective mass plot for $P$-wave at $Ma=1.0$~.

\end{enumerate}

\vfill
\pagebreak

\section*{Tables}
\vskip 5 truemm
{}~~~~~~~~~~~~~~~~~~~~~~~~~~~~~~~~~~~~~

{\bf Table 1.} \vskip 10 truemm
\begin{tabular}{||l|l|l||}          \hline
  State    &   $~~M_0a$    & $~~M_1a$ \\ \hline \hline
   $S$       &   1.098(2)  & 0.32(3) \\ \hline
   $P$       &   1.230(10)  & 0.29(4) \\ \hline
   $D1$      &   1.442(19)  & 0.66(22) \\ \hline
   $D2$      &   1.415(31)  & 0.68(22) \\ \hline
   $H1$      &   1.42(11) & ~~-  \\ \hline
   $H2$      &   1.37(8)  & ~~-  \\ \hline
\end{tabular}
\vspace{10mm}

{\bf Table 2.} \vskip 10 truemm
\begin{tabular}{||l|l|l|l||}          \hline
 $ Ma$     & $ M_{S}a $ & $ M_{P}a $ &  $S/P$-split    \\ \hline \hline
   1.0     &  1.38(1)  & 1.51(6) &  0.13(7)  \\ \hline
   1.3     &  1.20(1)  & 1.32(2) &  0.12(3)  \\ \hline
   1.6     &  1.086(9) & 1.22(3) &  0.14(4)  \\ \hline
   1.9     &  1.003(8) & 1.14(2) &  0.14(2)  \\ \hline
\end{tabular}

\vfill

\end{document}